\documentclass[twocolumn,superscriptaddress,showpacs,aps,prb]{revtex4}

\usepackage{amssymb,amsmath}
\usepackage[dvips]{graphicx}
\usepackage[footnotes,writekey=false]{notes2bib}

\usepackage{hyperref}
\hypersetup{
breaklinks=true,
colorlinks=true,
citecolor=blue,
linkcolor=blue,
filecolor=blue,
urlcolor=blue
}

\begin{document}

\title{Infinite-randomness quantum critical points induced by dissipation}

\author{Thomas Vojta}
\affiliation{Department of Physics, Missouri University of Science and Technology, Rolla,
MO 65409, USA}
\author{Chetan Kotabage}
\affiliation{Department of Physics, Missouri University of Science and Technology, Rolla,
MO 65409, USA}
\author{Jos\'{e} A. Hoyos}
\affiliation{Department of Physics, Missouri University of Science and Technology, Rolla,
MO 65409, USA} \affiliation{Department of Physics, Duke University, Durham, NC 27708,
USA}

\begin{abstract}
We develop a strong-disorder renormalization group to study quantum phase transitions
with continuous $O(N)$ symmetry order parameters under the influence of both quenched
disorder and dissipation. For Ohmic dissipation, as realized in Hertz' theory of the
itinerant antiferromagnetic transition or in the superconductor-metal transition in
nanowires, we find the transition to be governed by an exotic infinite-randomness fixed
point in the same universality class as the (dissipationless) random transverse-field
Ising model. We determine the critical behavior and calculate key observables at the
transition and in the associated quantum Griffiths phase. We also briefly discuss the
cases of superohmic and subohmic dissipation.
\end{abstract}


\date{\today}
\maketitle

\section{Introduction}

In recent years, it has become clear that quenched disorder, i.e., impurities, defects or
other types of imperfections can significantly modify the low-temperature behavior of
quantum many-particle systems. At zero-temperature quantum phase transitions, the
interplay between large-scale quantum fluctuations and random fluctuations leads to much
more dramatic disorder effects than at classical thermal phase transitions, resulting in
various exotic phenomena such as quantum Griffiths effects
\cite{ThillHuse95,GuoBhattHuse96,RiegerYoung96}, non-power-law dynamical scaling
\cite{Fisher92,Fisher95}, or even smeared phase transitions. \cite{Vojta03a,HoyosVojta08}
A recent review of some of these phenomena can be found in Ref.\ \onlinecite{Vojta06}.

The quantum phase transitions in random transverse-field Ising magnets are among the most
striking examples of such behavior. Utilizing a real-space renormalization group (RG)
technique due to Ma, Dasgupta and Hu,\cite{MaDasguptaHu79,DasguptaMa80} now known as the
strong-disorder RG (see Ref.\ \onlinecite{IgloiMonthus05} for a review), Fisher
\cite{Fisher92,Fisher95} showed that the one-dimensional random transverse-field Ising
chain features an unconventional infinite-randomness critical point with ultraslow
activated rather than power-law dynamical scaling. It is accompanied by strong power-law
quantum Griffiths effects in the vicinity of the transition. While it was initially
suspected that this scenario is special to one space dimension, Motrunich \emph{et al.}\
\cite{MMHF00} showed that the random transverse-field Ising models in two and three
dimensions also display infinite-randomness critical points.

A dissipative environment further hampers the dynamics. In the experimentally relevant
case of Ohmic damping, the large locally ordered droplets that are normally responsible
for quantum Griffiths effects completely cease to tunnel.
\cite{CastroNetoJones00,MillisMorrSchmalian01,MillisMorrSchmalian02} Instead, they
develop static order independently from the bulk system which destroys the sharp quantum
phase transition by smearing. \cite{Vojta03a,HoyosVojta08} A similar freezing of locally
ordered droplets also occurs close to a quantum percolation transition.
\cite{HoyosVojta06,HoyosVojta08b}

The above behavior of Ising order parameters must be contrasted with that of
\emph{continuous} $O(N)$ symmetry order parameters. While the bulk ground state phases of
one-dimensional Heisenberg random quantum spin chains are governed by infinite-randomness
fixed points,\cite{MaDasguptaHu79,DasguptaMa80,Fisher94} higher-dimensional random
quantum Heisenberg systems have more conventional ground states,\cite{LMRI03,LRLI06} and
their quantum phase transitions are governed by conventional critical points.
\cite{Sandvik02,VajkGreven02,SknepnekVojtaVojta04,VojtaSknepnek06} As in the Ising case,
adding Ohmic dissipation hampers the dynamics of $O(N)$ symmetric order parameters. Vojta
and Schmalian \cite{VojtaSchmalian05} showed that the ``energy gap'' of large locally
ordered droplets is exponentially small in their volume leading to power-law quantum
Griffiths effects analogous to those in the \emph{dissipationless} random
transverse-field Ising model. This analogy suggests the important question of whether
Ohmic dissipation can induce an unconventional infinite-randomness quantum critical point
for a continuous $O(N)$ symmetry order parameter.

In addition to its significance for the theory of quantum phase transitions,
this question also has important experimental applications. Consider the
antiferromagnetic quantum phase transition of itinerant electrons (as observed, e.g.,
in heavy fermion compounds \cite{Stewart01,Stewart06}). Within the
standard Hertz-Millis spin-fluctuation theory \cite{Hertz76,Millis93}, it is described by
an $O(3)$ Landau-Ginzburg-Wilson (LGW) order parameter field theory with Ohmic dissipation.
The properties of this transition have been a long-standing unsolved problem; and given
the fact that most experimental systems are rather dirty, studying the effects of disorder
on the Hertz-Millis theory is of prime interest.

A second potential application is provided by the pair breaking superconductor-metal
quantum phase transitions in nanowires. \cite{RogachevBezryadin03} It can be described by
a one-dimensional $O(2)$ LGW theory with Ohmic dissipation.
\cite{SachdevWernerTroyer04,DRSS08,DelMaestroRosenowSachdev08} There is experimental
evidence that the pair breaking in this systems is caused by surface magnetic impurities
which necessarily also introduce quenched disorder.

In this paper, we investigate the quantum phase transition of a continuous-symmetry
$O(N)$ order parameter under the combined influences of both quenched disorder and Ohmic
dissipation. To this end we develop a strong-disorder RG suitable for
this problem. The paper is organized as follows: In Sec.\ \ref{sec:LGW} we introduce
our model, a dissipative LGW order parameter field theory. In Sec.\ \ref{sec:SDRG}
we implement the strong-disorder RG in the large-$N$ limit
and relate it to that of the random transverse-field
Ising model. We also summarize the solution and the resulting critical behavior.
In Sec.\ \ref{sec:Observables}, we calculate key observables close to the transition
while Sec.\ \ref{sec:Generalizations} deals with the case of non-Ohmic damping.
We also show that our results do not rely on the large-$N$ limit.
We conclude in Sec.\ \ref{sec:Conclusions}.

A short account of part of this work has already been published in Ref.\
\onlinecite{HoyosKotabageVojta07}.

\section{Order parameter field theory}
\label{sec:LGW}

We start from a quantum Landau-Ginzburg-Wilson (LGW) free energy functional for an
$N$-component ($N>1$) vector order parameter $\varphi=(\varphi_1,\dots,\varphi_N)$ in $d$
space dimensions. For the above-mentioned itinerant antiferromagnetic quantum phase
transition, generically $d=3$ and $N=3$ while for the superconductor-metal transition in
nanowires, $d=1$ and $N=2$. The LGW free energy can be derived from
an appropriate microscopic Hamiltonian of disordered electrons using standard methods
\cite{Hertz76,Millis93,KirkpatrickBelitz96}
(for a critical discussion of this approach see Ref.\ \onlinecite{BelitzKirkpatrickVojta05}).
In the absence of quenched disorder, the action of our LGW theory reads
\begin{equation}
S=\int{\rm d}y{\rm d}x ~\varphi(x)\Gamma(x,y)\varphi(y)+\frac{u}{2N}\int{\rm d}x~\varphi^{4}(x)\,,
\label{eq:clean-action}
\end{equation}
where $x\equiv(\mathbf{x},\tau)$ comprises imaginary time $\tau$ and position
$\mathbf{x}$, $\int{\rm d}x\equiv\int{\rm d}\mathbf{x}\int_{0}^{1/T}{\rm d}\tau$, and $u$
is the standard quartic coefficient. $\Gamma(x,y)$ denotes the bare inverse propagator
(two-point vertex) whose Fourier transform reads
\begin{equation}
\Gamma(\mathbf{q},\omega_{n})=r+\xi_{0}^{2}\mathbf{q}^{2}+\gamma_0\left|\omega_{n}\right|^{2/z_{0}}~.
\label{eq:bare_Gamma}
\end{equation}
Here, $r$ is the bare distance from criticality (the bare gap), $\xi_{0}$ is a
microscopic length scale, and $\omega_{n}$ is a Matsubara frequency. The nonanalytic
frequency dependence of $\Gamma(\mathbf{q},\omega_{n})$ is caused by the coupling of the
order parameter to a dissipative bath. We are mostly interested in the case of overdamped
(Ohmic) dynamics corresponding to a value of $z_0=2$. However, to demonstrate the special
role of $z_{0}=2$, we also consider different values of $z_{0}$. The damping coefficient
$\gamma_0$ depends on the coupling of the order parameter to the dissipative bath and the
spectral density of the bath modes.

In the presence of quenched disorder, the functional form of the order-parameter field
theory (\ref{eq:clean-action}) does not change qualitatively, but the distance from criticality
$r$ becomes a random function of spatial position. Analogously, disorder appears in
$\xi_0,\gamma_0$ and $u$.

Let us briefly comment on possible modifications of the two-point vertex
(\ref{eq:bare_Gamma}) by mode-coupling effects. For the itinerant ferromagnetic quantum
phase transition \cite{KirkpatrickBelitz96b,VBNK96,BelitzKirkpatrickVojta97} and the
superconductor-metal transition without magnetic impurities,
\cite{KirkpatrickBelitz97,SknepnekVojtaNarayanan04} the coupling between the order
parameter fluctuations and the soft particle-hole excitations of the metal leads to a
long-range interaction in space, represented by a nonanalytic $q$-dependence instead of
the simple $\mathbf{q}^2$ term. In contrast, in our examples, the $\mathbf{q}^2$ term
remains the leading term because the relevant modes are either gapped (for the
superconductor-metal transition due to magnetic impurities) or couple too weakly to the
order parameter (in the case of the itinerant antiferromagnetic transition).
\cite{BelitzKirkpatrick94}

Our goal is the application of the real-space-based strong-disorder RG.
We therefore need to discretize the continuum action (\ref{eq:clean-action}) in
space (but not in imaginary time) by defining discrete coordinates $\mathbf{x}_j$ and rotor
variables $\varphi_j(\tau)$. These rotors are coarse-grained rather than microscopic variables,
they represent the average order parameter in a volume $\Delta V$ large compared to the
microscopic scale $\xi_0$ but small compared to the true correlation length $\xi$, i.e.,
$\varphi_j(\tau) = \int_{\Delta V} d\mathbf{y} \varphi(\mathbf{x}_j +\mathbf{y},\tau)$.

For simplicity, we first consider the large-$N$ limit of our LGW theory. This will allow us
to perform all of the following calculations explicitly. We will later show that the
RG fixed point is the same for all $N>1$. The resulting discrete
large-$N$ action reads
\begin{eqnarray}
S & = & T \sum_{i}\sum_{\omega_{n}}\left(r_{i}+\lambda_{i}+\gamma_{i}\left|\omega_{n}\right|^{2/z_{0}}\right)\left|\phi_{i}(\omega_{n})\right|^{2}\nonumber \\
 &  & -T \sum_{\left\langle i,j\right\rangle }\sum_{\omega_{n}}\phi_{i}(-\omega_{n})J_{ij}\phi_{j}(\omega_{n})\,,
\label{eq:action}
\end{eqnarray}
where $r_{i}$, $\gamma_{i}>0$ and the nearest-neighbor interactions $J_{ij}>0$ are random
quantities, and
$\phi_{j}(\omega_{n})=\int_{0}^{1/T}\varphi_{j}(\tau)e^{i\omega_{n}\tau}{\rm d}\tau$ is
the Fourier transform of the rotor variable. The Lagrange multipliers $\lambda_{i}$
enforce the large-$N$ constraints $\langle(\varphi_{i}^{(k)}(\tau))^{2}\rangle=1$ for
each order parameter component $\varphi_{i}^{(k)}$ at each site $i$; they have to be
determined self-consistently. The renormalized local distance from criticality at site
$i$ is given by $\epsilon_{i}=r_{i}+\lambda_{i}$. In the disordered phase, \emph{all}
$\epsilon_i>0$.

\section{Strong-disorder renormalization group}
\label{sec:SDRG}

The idea of the strong-disorder (Ma-Dasgupta-Hu) RG
\cite{MaDasguptaHu79,DasguptaMa80,Fisher92,Fisher95} consists in the successive
decimation of local high-energy degrees of freedom. It relies on the disorder
distributions being broad and becomes exact in the limit of infinitely broad
distributions. For now, we assume that our distributions are sufficiently broad, we will
later show that their widths diverge at the critical RG fixed point, justifying the
method.

\subsection{Single-cluster solution}
\label{subsec:single_cluster}

Let us start by considering a single rotor variable $\varphi$ (i.e., a single cluster)
with the action
\begin{equation}
S_{\rm cl} = T \sum_{\omega_n}\left(r+\lambda + \gamma|\omega_n|^{2/z_0} \right)
|\phi(\omega_n)|^2~. \label{eq:single-site-action}
\end{equation}
The value of the Lagrange multiplier $\lambda$ is determined by the length constraint
\begin{equation}
1 = \langle \phi^2 \rangle = T \sum_{\omega_n} \frac 1 {r +\lambda +\gamma
|\omega_n|^{2/z_0} }~. \label{eq:single-site-constraint}
\end{equation}
At zero temperature, the Matsubara sum can be turned into an integral, resulting in
\begin{equation}
1= \frac 1 {2\pi} \int_{-\infty}^{\infty} d\omega \frac 1 {\epsilon + \gamma
|\omega_n|^{2/z_0}}~, \label{eq:single-site-constraint-integral}
\end{equation}
where $\epsilon=r + \lambda$ is the renormalized distance from criticality.

To proceed, we now need to distinguish super-Ohmic, Ohmic and sub-Ohmic dissipation. In
the super-Ohmic case, $z_0<2$, the integral can be carried out straight-forwardly giving
$\langle \phi^2 \rangle= c \epsilon^{(z_0-2)/2}\gamma^{-z_0/2}$ with $c$ being a
constant. Solving for $\epsilon$ yields the relation between the gap and the damping
constant (i.e., the cluster size)
\begin{equation}
\epsilon \sim (1/\gamma)^{z_0/(2-z_0)}~. \label{eq:gap-super-Ohmic}
\end{equation}
In contrast, we need to introduce a high-frequency cutoff $\Lambda$ to carry out the
constraint integral in the Ohmic case $z_0=2$, giving $\langle \phi^2
\rangle=\ln(1+\gamma\Lambda/\epsilon)/\pi\gamma$. The resulting dependence of the gap on
the damping constant is exponential,
\begin{equation}
\epsilon = \gamma \Lambda /\left(e^{\pi \gamma}-1\right) \approx \gamma \Lambda e^{-\pi
\gamma}~, \label{eq:gap_Ohmic}
\end{equation}
signifying that a single Ohmic cluster is marginal, i.e., right at the lower critical
``dimension'' of the problem. In the sub-Ohmic case, $z_0>2$, the single-cluster physics
changes dramatically: The constraint integral in
(\ref{eq:single-site-constraint-integral}) converges in the limit $\epsilon\to 0$. Thus,
once $\gamma > \gamma_c=\Lambda^{(z_0-2)/z_0} z_0/(\pi(z_0-2))$, Eq.\
(\ref{eq:single-site-constraint-integral}) does not have a solution $\epsilon>0$ anymore,
implying that the rotor has undergone a localization phase transition caused by the
sub-Ohmic dissipation.

\subsection{Recursion relations}
\label{subsec:recursions}

In our large-$N$ action (\ref{eq:action}), the competing independent local energies are
the gaps $\epsilon_i$ and the interactions $J_{ij}$ (the damping coefficient $\gamma_i$
and the gap $\epsilon_i$ are not independent, they are coupled via the large-$N$
constraint at site $i$). In the bare theory, the $J_{ij}$ and $\epsilon_i$ are
independent random variables with probability distributions $P(J)$ and $R(\epsilon)$,
respectively. Each step of the strong disorder RG eliminates one rotor variable by first
identifying the largest local energy $\Omega=\max(\epsilon_i,J_{ij})$ and then decimating
the associated high-energy degree of freedom.

\subsubsection{Decimating a site}

Specifically, if the largest local energy is a gap, say $\epsilon_2$, the corresponding
rotor $\phi_2$ is far away from criticality and does not contribute to the macroscopic
order parameter. However, integrating out its fluctuations generates effective
interactions between all pairs of sites that couple to $\phi_2$. If the disorder
distributions are broad, $\epsilon_2$ is much larger than all local energies associated
with the neighboring sites. Thus, $\phi_2$ can be integrated out in perturbation theory
with the unperturbed part of the action being
\begin{equation}
S_{0}=T\sum_{\omega_{n}}(\epsilon_{2}+\gamma_{2}\left|\omega_{n}\right|^{2/z_{0}})|\phi_{2}(\omega_{n})|^{2}
\end{equation}
while $S_1=S-S_0$ is the perturbation. Up to 2nd order in perturbation theory, we only
need to consider the interaction of $\phi_2$ with the neighboring sites $j$, thus
\begin{equation}
S_{1}=-T\sum_{j\ne 2,\omega_{n}}J_{2j}\phi_{2}(-\omega_{n})\phi_{j}(\omega_{n})~.
\label{eq:action-S1}
\end{equation}
The partition function can now be written as
\begin{eqnarray}
Z&=&\int D[\phi_2]\prod_{j\ne 2} D[\phi_j] e^{-S} = Z_0 \int \prod_{j\ne 2} D[\phi_j]
\langle e^{-S_1} \rangle_0 \nonumber \\
 &=& Z_0 \int \prod_{j\ne 2} D[\phi_j] e^{-\tilde S}~,
\label{eq:partition-function}
\end{eqnarray}
where $\int D[\phi_j]$ comprises integration over all frequency components of $\phi_j$,
$Z_0$ is the partition function associated with the action $S_0$, and $\langle \cdot
\rangle_0$ denotes the average w.r.t. $S_0$. The renormalized action $\tilde S$ can be
calculated in cumulant expansion
\begin{equation}
\tilde S = -\ln \langle e^{-S_1} \rangle_0 = \langle S_1 \rangle_0 - \frac 1 2 \left[
\langle S_1^2 \rangle_0 - \langle S_1 \rangle_0^2 \right] \pm \ldots ~.
\label{eq:cumulant}
\end{equation}
Evaluating the averages, we obtain $\langle S_1 \rangle_0=0$ due to symmetry and
\begin{eqnarray}
\langle S_1^2 \rangle_0 = T \sum_{\omega_n} \left[ \sum_j \frac {J_{2j}^2
|\phi_j(\omega_n)|^2} {\epsilon_2 + \gamma_2 |\omega_n|^{2/z_0}} \right. \qquad \nonumber
\\ \qquad \left . + 2 \sum_{i\ne j} \frac {J_{i2}J_{2j}
\phi_i^*(\omega_n)\phi_j(\omega_n)} {\epsilon_2 + \gamma_2 |\omega_n|^{2/z_0}} \right] ~.
\label{eq:S1-squared}
\end{eqnarray}
The first term in the square brackets just gives subleading renormalizations of the gaps
$\epsilon_j$ of the neighboring sites and can thus be dropped. The second term provides
the renormalized interactions $\tilde J_{ij}$ between all sites that used to couple to
$\phi_2$. Their leading low frequency behavior is $\tilde J_{ij} =
J_{i2}J_{2j}/\epsilon_2$ independent of the exponent $z_0$. This term has to be added to
the interaction $J_{ij}$ already coupling sites $i$ and $j$, if any. Consequently, the
final recursion relation for the RG step reads
\begin{equation}
\tilde{J}_{ij}=J_{ij}+\frac{J_{i2}J_{2j}}{\epsilon_{2}}\,.
\label{eq:J-tilde}
\end{equation}
At the end of the RG step, $\phi_{2}$ is dropped from the action. Note that the
multiplicative structure of the effective interaction in (\ref{eq:J-tilde}) is a direct
consequence of second order perturbation theory. It does not depend on details of the
model, in particular, it is valid for \emph{any} $z_0$.

\subsubsection{Decimating a bond}

Let us now consider the RG step in the case of the largest local energy being an
interaction, say $J_{23}$ coupling sites 2 and 3. For broad disorder distributions,
$J_{23} \gg \epsilon_2,\epsilon_3$. Thus, the two rotors $\phi_2$ and $\phi_3$ are
essentially parallel and can be replaced by a single rotor $\tilde \phi_2$ which
represents the entire cluster comprising $\phi_2$ and $\phi_3$. The moment $\tilde \mu_2$
of the effective rotor, i.e., the number of original sites in the cluster is the sum of
the moments $\mu_2$ and $\mu_3$ of the original rotors,
\begin{equation}
\tilde{\mu}_{2}=\mu_{2}+\mu_{3}\,.
\label{eq:moment}
\end{equation}
To find the renormalized gap $\tilde \epsilon_2$ of the effective rotor, we solve exactly
the two-site problem involving $\phi_2$ and $\phi_3$ while treating the couplings to all
other sites as perturbations. The two-site action is given by
\begin{eqnarray}
S_{0} &=& ~T\sum_{\omega_{n}}\sum_{i=2,3}(r_i+\lambda_i+\gamma_{i}\left|
\omega_{n}\right|^{2/z_{0}})|\phi_{i}(\omega_{n})|^{2} \nonumber \\ &&-
T\sum_{\omega_{n}}J_{23}\phi_{2}(-\omega_{n})\phi_{3}(\omega_{n})~.
\label{eq:two-site-action}
\end{eqnarray}
It is subject to the large-$N$ length constraints
\begin{eqnarray}
1&=& \langle \phi_2^2 \rangle  = T\sum_{\omega_n}  \frac {d_3}{d_2 d_3-J^2/4}~,
    \nonumber \\
1&=& \langle \phi_3^2 \rangle  = T\sum_{\omega_n} \frac {d_2}{d_2 d_3-J^2/4}~,
\label{eq:integral-equations}
\end{eqnarray}
with $d_j = r_j +\lambda_j +\gamma_j |\omega_n|^{2/z_0}$. They determine the Lagrange
multipliers $\lambda_i$. (It is important to note that the value of $r_i+\lambda_i$ in
the two-site cluster is different from the single-site $\epsilon_i$.)

To integrate out the high-energy mode, we diagonalize the quadratic form in
(\ref{eq:two-site-action}) separately for each Matsubara frequency. The two eigenvalues
read
\begin{eqnarray}
\kappa_{a,b} &=& \frac 1 2 \left( d_2+d_3 \pm \sqrt{(d_2-d_3)^2 +J_{23}^2} \right)
\nonumber
\\
 &=& \frac 1 2 \left( d_2 +d_3 \pm J_{23} \right) +O\left(\frac {\epsilon_j}{J_{23}},\frac
 {\omega_n}{J_{23}}\right)~.~
\label{eq:eigenvalues}
\end{eqnarray}
The corresponding eigenmodes are given by $\psi_a=\alpha \phi_2 +\beta \phi_3$ and
$\psi_b = -\beta\phi_2 +\alpha \phi_3$ with
\begin{eqnarray}
\alpha &=& \frac
{d_3-d_2+\sqrt{(d_2-d_3)^2+J^2}}{\sqrt{(d_3-d_2+\sqrt{(d_2-d_3)^2+J^2})^2 +J^2}}~,
\nonumber \\ \beta  &=& \frac {J}{\sqrt{(d_3-d_2+\sqrt{(d_2-d_3)^2+J^2})^2 +J^2}}~.~
\label{eq:coefficients}
\end{eqnarray}
The higher eigenvalue $\kappa_b$ is at least $J_{23}$ above the lower eigenvalue
$\kappa_a$; we thus integrate out the corresponding mode leaving us with the effective
action $\tilde S = T \sum_{\omega_n} \lambda_a |\psi_a(\omega_n)|^2$ and a length
constraint $\langle \psi_a^2 \rangle = \langle (\alpha \phi_2 +\beta \phi_3)^2\rangle \ne
1$.

We define the renormalized rotor variable by rescaling $\tilde \phi_2 = \psi_a /\langle
\psi_a^2\rangle^{1/2}$ because we wish it to fulfill the same length constraint $\langle
\tilde \phi_2^2 \rangle =1$ as all other rotor variables. Inserting this definition into
the (diagonalized) two-site action (\ref{eq:two-site-action}) allows us to identify the
renormalized gap, damping constant, and interactions with the neighbors.
\begin{eqnarray}
{\tilde \epsilon}_2 &=&\frac 1 2 \langle \psi_a^2 \rangle (r_2 +\lambda_2 +r_3
+\lambda_3-J)~, \nonumber \\ {\tilde \gamma}_2 &=&\frac 1 2 \langle \psi_a^2 \rangle
(\gamma_2 + \gamma_3)~, \nonumber \\ {\tilde J}_{2j} &=& \langle \psi_a^2 \rangle^{1/2}
(\alpha J_{2j} + \beta J_{3j})|_{\omega_n \to 0}~. \label{eq:renormalized_parameters}
\end{eqnarray}
To proceed further, we need explicit results for the Lagrange multipliers $\lambda_2$ and
$\lambda_3$ as well as $\langle \psi_a^2\rangle$. This requires the solution of the two
coupled integral equations (\ref{eq:integral-equations}). In the case of Ohmic
dissipation, the integrals are rational and can be done exactly. In the limit of strong
disorder, we obtain $r_2 +\lambda_2 +r_3 +\lambda_3-J = 2\epsilon_2\epsilon_3/J$ and
$\alpha|_{\omega_n \to 0}=\beta|_{\omega_n \to 0}=\sqrt{2}/2$.  Moreover, $\langle
\psi_a^2 \rangle$ is bounded between 1 and 2 and approaches 2 in the asymptotic limit
$\Omega \to  0$. This leads to the recursion relations
\begin{equation}
\tilde{J}_{2j}=J_{2j}+J_{3j}~, \label{eq:J-tilde-cluster}
\end{equation}
\begin{equation}
\tilde{\epsilon}_{2}=2\frac{\epsilon_{2}\epsilon_{3}}{J_{23}}\,, \label{eq:e-tilde}
\end{equation}
implying an additive relation for the renormalized damping constant,
\begin{equation}
\tilde\gamma_{2}=\gamma_{2}+\gamma_{3}~.
\label{eq:gamma-tilde}
\end{equation}
We emphasize that the multiplicative form of (\ref{eq:e-tilde}) is \emph{not} independent
of the functional form of the action (\ref{eq:action}). In contrast to the recursion
relation (\ref{eq:J-tilde}) for the interactions, the recursion relation
(\ref{eq:e-tilde}) for the gaps is special to the case of Ohmic dissipation. It is
related to the fact that the gap $\epsilon$ of single cluster depends exponentially on
the damping constant (and thus on the cluster size), $\epsilon = \gamma \Lambda e^{-\pi
\gamma}$, as derived in (\ref{eq:gap_Ohmic}). We will come back to this point in Sec.\
\ref{sec:Generalizations} where we discuss the case of non-Ohmic dissipation.

Although the prefactor in Eq.\ (\ref{eq:e-tilde}) is larger than 1, this does \emph{not}
mean that the renormalized gap can become larger than the decimated ones in the weak
disorder limit.  Using the methods of Ref.\ \onlinecite{Hoyos08} we showed that the exact
value (\ref{eq:renormalized_parameters}) of $\tilde{\epsilon}_2$ [calculated within the
two-site action (\ref{eq:two-site-action})] is always less than the decimated gaps
($\epsilon_2$ and $\epsilon_3$) for all $\epsilon_2 , \epsilon_3 \leq J_2$. Therefore,
the system flows towards the infinite-randomness fixed point for all bare disorder
strengths, ensuring the internal consistency of the RG.

The net result of a single RG step is the elimination of one rotor and the reduction of
the maximum local energy $\Omega$ as well as renormalizations of the remaining energies
and reconnections of the lattice.

The RG recursion relations (\ref{eq:J-tilde}), (\ref{eq:moment}),
(\ref{eq:J-tilde-cluster}) and (\ref{eq:e-tilde}) completely define the RG procedure.
They are identical to the corresponding relations for the \emph{dissipationless} random
transverse-field Ising model. \cite{Fisher92,Fisher95,MMHF00} We thus conclude that our
system belongs to the same universality class. Note, however, that there are some subtle
differences in the behavior of some observables due to the continuous symmetry of the
order parameter and the Ohmic damping, as will be discussed in Section
\ref{sec:Observables}.

\subsection{RG flow equations and fixed points}
\label{subsec:IRFP}

In this subsection, we briefly summarize Fisher's solution
\cite{Fisher92,Fisher95} of the strong-disorder RG defined by the recursions
 (\ref{eq:J-tilde}), (\ref{eq:moment}),
(\ref{eq:J-tilde-cluster}) and (\ref{eq:e-tilde}) to the extent necessary for the purposes of this
paper.

In one space dimension, the RG step does not change the lattice topology because the
interactions remain between nearest-neighbor sites only, and the $\epsilon$ and $J$
remain statistically independent. Therefore, the theory can be formulated in terms of the
probability distributions $P(J)$ and $R(\epsilon)$. Fisher derived RG flow equations for
these distributions and solved them analytically. There are three types of nontrivial
fixed points corresponding to the ordered and disordered quantum Griffiths phases and the
quantum critical point that separates them.

The most remarkable feature of the critical fixed point is that the probability
distributions $P$ and $R$ broaden without limit under renormalization, even on a
logarithmic scale. Using logarithmic variables $\Gamma=\ln(\Omega_{I}/\Omega)$ {[}where
$\Omega_{I}$ is of the order of the initial (bare) value of $\Omega${]},
$\zeta=\ln(\Omega/J)$ and $\beta=\ln(\Omega/\epsilon)$ the probability distributions
$\cal P(\zeta)$ and $\cal R(\beta)$ at the critical fixed point read
\begin{eqnarray}
{\cal P} (\zeta) = \frac 1 \Gamma e^{-\zeta/\Gamma}, \quad  {\cal R} (\beta) = \frac 1 \Gamma
e^{-\beta/\Gamma}~.
\label{eq:critical-FP}
\end{eqnarray}
The diverging widths of the probability distributions give the critical point its name,
\emph{viz.} infinite-randomness critical point. They also \emph{a posteriori} justify the
method, because the perturbative recursion relations
(\ref{eq:J-tilde})--(\ref{eq:e-tilde}) become exact in the limit of infinitely broad
distributions (i.e., approaching the critical point).

The complete critical behavior can be found by including the moments and lengths of the
clusters in the RG procedure. It is characterized by three exponents $\nu=2$, $\psi=1/2$,
and $\phi=(1+\sqrt{5})/2$. The correlation length exponent $\nu$ determines how the
average correlation length $\xi$ diverges if one approaches the critical point via
\begin{equation}
\xi \sim |r|^{-\nu}~.
\label{eq:exponent_nu}
\end{equation}
Here $r$ denotes the fully renormalized dimensionless distance from criticality which is
given by $r \sim [\ln(\epsilon)] -[\ln(J)]$ in terms of the bare variables ($[\cdot]$
denotes the disorder average).\footnote{In principle, one must distinguish between the
bare distance from criticality appearing in (\ref{eq:bare_Gamma}) and the renormalized one
appearing in the scaling relations. We will suppress this difference unless it is of
importance for our considerations.}

The tunneling exponent $\psi$ controls the dynamical scaling, i.e., the relation between
length scale $L$ and energy scale $\Omega$. It is of activated rather than power-law type
\begin{equation}
\ln(\Omega_I/\Omega) \sim L^\psi~,
\label{eq:exponent_psi}
\end{equation}
which is a direct consequence of the multiplicative structure of the recursions
(\ref{eq:J-tilde}) and (\ref{eq:e-tilde}). $\psi$ also controls the density $n_\Omega$
of clusters surviving at an energy scale $\Omega$ in the RG procedure. Its scaling form
is given by
\begin{equation}
n_\Omega(r) = [\ln(\Omega_I/\Omega)]^{-d/\psi}
X_n\left[r^{\nu\psi}\ln(\Omega_I/\Omega)\right]~,
\label{eq:n_scaling}
\end{equation}
with the scaling function behaving as $X_n(0) =$ const and $X_n(y\to\infty)
\sim y^{d/\psi}e^{-cdy}$ where $c$ is a constant. As a result, the cluster density
decreases as  $n_\Omega \sim[\ln(\Omega_I/\Omega)]^{-d/\psi}$ at criticality while it
behaves as $n_\Omega \sim r^{d\nu} \Omega^{d/z}$ in the disordered quantum Griffiths
phase ($r>0$). The dynamical exponent $z$ varies with $z\sim r^{-\nu\psi}$ in the
Griffiths phase.

Finally, the exponent $\phi$ describes how the typical moment $\mu_\Omega$ of a surviving
cluster depends on the energy scale $\Omega$. The scaling form of $\mu_\Omega$ reads
\begin{equation}
\mu_\Omega(r) = [\ln(\Omega_I/\Omega)]^{\phi}
X_\mu\left[r^{\nu\psi}\ln(\Omega_I/\Omega)\right]~.
\label{eq:mu_scaling}
\end{equation}
The scaling function behaves as $X_\mu(0) =$ const and $X_\mu(y\to\infty)
\sim y^{1-\phi}$. Thus, at criticality the typical moment increases as
$\mu_\Omega \sim [\ln(\Omega_I/\Omega)]^{\phi}$ while it behaves as
$\mu_\Omega \sim r^{\nu\psi(1-\phi)} \ln(\Omega_I/\Omega)$
in the disordered quantum Griffiths phase.

The strong-disorder RG steps discussed in subsection \ref{subsec:recursions} generate
effective interactions between sites that were previously uncoupled. In dimensions $d>1$,
this changes the lattice connectivity, and it introduces statistical correlations between
the $J$ and $\epsilon$. Therefore, the theory cannot be formulated in terms of individual
probability distributions of these variables, and a closed-form analytical solution
appears to be impossible. However, Motrunich \emph{et al.} \cite{MMHF00} numerically
implemented the recursion relations (\ref{eq:J-tilde})--(\ref{eq:e-tilde}) in two
dimensions, keeping track of all reconnections of the lattice under the RG. They found an
infinite randomness critical point very similar to that in one dimension. In fact, the
critical behavior described in Eqs.\ (\ref{eq:exponent_nu})--(\ref{eq:mu_scaling}) is
also valid in two dimensions, but with different exponent values. Various numerical
implementations\cite{MMHF00,LKIR00,KLRKI01,LinIgloiRieger07} of the strong disorder RG
yielded $\psi=0.42\ldots0.6$, $\phi=1.7\ldots2.5$ and $\nu=1.07\ldots1.25$. In three
dimensions, the RG flow towards an infinite-randomness fixed point has been confirmed
\cite{MMHF00}, but reliable estimates of the exponent values are still missing.

The strong-disorder RG allows one to identify the infinite-randomness fixed point and
confirm its stability but, strictly, it cannot answer the question of whether or not a
weakly or moderately disordered system will flow towards this fixed point because, if the
disorder is weak, the strong disorder RG step is not very accurate. [An internal
consistency check \cite {Hoyos08}  of the RG in the weak disorder limit can be achieved
by computing exactly rather than perturbatively the renormalized couplings (gaps and
interactions) within the relevant two-site or three-site clusters, see Sec.\
\ref{subsec:recursions}.] For our system, additional insight can be gained from the
results of a conventional perturbative (replica based) renormalization group. Building on
earlier work, \cite{BoyanovskyCardy82} Kirkpatrick and Belitz \cite{KirkpatrickBelitz96}
showed that the perturbative RG always takes the system to large disorder strength even
if the bare disorder is very small. Moreover, by taking rare region effects into account
in an approximate way, Narayanan \emph{et al.}\cite{NVBK99a,NVBK99b} showed that there is
no stable weak-disorder fixed point; instead the perturbative RG shows runaway flow
towards large disorder. This strongly suggests that our infinite-randomness critical
point is universal and governs the quantum phase transition for all nonzero disorder
strength.

\section{Observables}
\label{sec:Observables}

The strategy for calculating, within the strong-disorder RG, thermodynamic observables
such as the susceptibility as a function of temperature consists in running the RG from
the initial energy scale $\Omega_I$ down to $\Omega=T$. The high-energy degrees of
freedom eliminated in this procedure generally do not make significant contributions to
the low-energy behavior of observables. At best, they change non-universal constants. All
clusters surviving at energy scale $\Omega=T$ can be considered to be independent because
they are coupled by interactions $J$ much smaller than $T$. The desired observable is
thus simply the sum of independent contributions from the individual surviving clusters.
Frequency-dependent observables can be determined analogously.

\subsection{Single-cluster results}

In order to proceed, we therefore need to calculate the relevant observables for single
clusters.
To do so, we add a source term to the single-cluster action (\ref{eq:single-site-action}). It reads
\begin{equation}
S_{H} = - T \sum_{\omega_n} H(\omega_n) \phi(-\omega_n)~,
\label{eq:source-term}
\end{equation}
with $H(\omega_n)=\int_{0}^{1/T} H(\tau)e^{i\omega_{n}\tau}{\rm d}\tau$ being the Fourier
transform of the source field conjugate to the order parameter. Because the theory
defined by $S_{\rm cl} + S_H$ is still Gaussian, the partition function $Z_H$ in the
presence of the field can be easily evaluated. The dynamic (Matsubara) susceptibility is
then given by
\begin{equation}
\chi(i\omega_n)=\frac 1 T \frac {\partial^2 \ln Z_H }{\partial H(\omega_n)
\partial H(-\omega_n)} = \frac  1 {r+\lambda+\gamma|\omega_n|^{2/z_0}}~.
\label{eq:susc-definition}
\end{equation}
For the temperature-dependent static susceptibility, we set $\omega_n=0$ and find the
distance from criticality, $\epsilon(T) = r +\lambda(T)$ as function of temperature. To
this end we solve the finite-temperature constraint equation
(\ref{eq:single-site-constraint}) yielding
\begin{equation}
\epsilon(T) = \left \{
\begin{array}{cc}
  \epsilon(0) + a T &\quad (\gamma T^{2/z0} \ll \epsilon(0)) \\
  T & \quad (\gamma T^{2/z0} \gg \epsilon(0))
\end{array}
\right .~,
\label{eq:epsilon-T}
\end{equation}
where $\epsilon(0)$ is the zero-temperature value determined by the constraint
integral (\ref{eq:single-site-constraint-integral}). In the super-Ohmic and Ohmic cases,
$\epsilon(0)$ is given by  Eqs.\ (\ref{eq:gap-super-Ohmic}) and (\ref{eq:gap_Ohmic}),
respectively. The constant $a$ is given by
$a=\pi\gamma$ in the Ohmic case and $a=2/(2-z_0)$ in the super-Ohmic case.

If the rotor variable $\phi$ represents a cluster of moment (number of sites) $\mu$,
its contribution to the uniform susceptibility is proportional to $\mu^2$ while
the contribution to the local susceptibility is proportional to $\mu$. By combining this
with (\ref{eq:susc-definition}) and (\ref{eq:epsilon-T}), we obtain
the uniform static order parameter susceptibility as a function of temperature
of a cluster of moment $\mu$ and distance $\epsilon$ from criticality,
\begin{equation}
\chi_{\rm cl}(T) = \left \{
\begin{array}{cc}
\mu^2/\epsilon & (\epsilon \gg T) \\
\mu^2/T        & (\epsilon \ll T)
\end{array} \right. ~.
\label{eq:cluster-chi}
\end{equation}
The corresponding results for the average \emph{local} susceptibility read
\begin{equation}
\chi_{\rm cl}^{\rm loc}(T) = \left \{
\begin{array}{cc}
 \mu/\epsilon & (\epsilon \gg T) \\
 \mu/T        & (\epsilon \ll T)
 \end{array} \right. ~.
\label{eq:cluster-chi_loc}
\end{equation}

To calculate the specific heat we also need the total energy contribution of a single
cluster which behaves as
\begin{equation}
\Delta E_{\rm cl} \sim  T  \quad (\epsilon \ll T).
\end{equation}
This is an important difference from the random transverse-field Ising model case\cite{Fisher95}
and stems from the fact that our rotor variables have an unbounded spectrum.

We now turn to the dynamical order parameter susceptibility at zero temperature
(focusing on Ohmic dissipation). From (\ref{eq:susc-definition}), we obtain
in imaginary time formalism
\begin{equation}
\chi_{\rm cl}(i\omega_n) = \frac {\mu^2} {\epsilon + \gamma|\omega_n|}~.
\end{equation}
After Wick rotation $i\omega_n \to \omega +i0$ to real frequencies, this leads
to  $\chi_{\rm
cl}(\omega+i0) = \mu^2/(\epsilon-i\gamma\omega)$ implying
\begin{equation}
{\rm Im} \chi_{\rm cl}(\omega+i0) = \frac {\mu^2 \, \gamma\omega} {\epsilon^2 +\gamma^2
\omega^2}~.
\label{eq:cluster-chi_dyn}
\end{equation}
Analogously, the dynamical local susceptibility reads
\begin{equation}
{\rm Im} \chi_{\rm cl}^{\rm loc}(\omega+i0) = \frac {\mu \, \gamma\omega} {\epsilon^2 +\gamma^2
\omega^2}~.
\label{eq:cluster-chi_dyn_loc}
\end{equation}

\subsection{Summing over all clusters}

We now combine the single-cluster observables summarized in the last subsection with the
strong-disorder RG results for density and moment of the surviving clusters given in
(\ref{eq:n_scaling}) and (\ref{eq:mu_scaling}). We focus on the critical point and the
disordered Griffiths phase. On the ordered side of the transitions, the scenario is
dimensionality-dependent because in $d>1$, an infinite percolating RG cluster forms
already at a finite energy scale. \cite{MMHF00}

To obtain the uniform static order
parameter susceptibility  $\chi(r,T)$ and the corresponding local susceptibility  $\chi^{\rm loc}(r,T)$,
we run the RG to the energy scale $\Omega=T$ and sum over all surviving clusters. Using
(\ref{eq:n_scaling}), (\ref{eq:mu_scaling}) and (\ref{eq:cluster-chi}), we obtain the scaling form
\begin{eqnarray}
\chi(r,T) &=& \frac 1 T n_T(r) \mu_T^2(r) \nonumber\\
          &=&
          \frac{1}{T}\left[\ln(\Omega_I/T)\right]^{2\phi-d/\psi}\Theta_{\chi}\left[r^{\nu\psi}\ln(\Omega_I/T)\right]~.~~
\label{eq:chi-scaling}
\end{eqnarray}
with the scaling function $\Theta_\chi$ given by $\Theta_\chi(y)=X_n(y) X_\mu^2(y)$.
At criticality, $r=0$, this leads to $\chi \sim
\left[\ln(\Omega_I/T)\right]^{2\phi-d/\psi}/T$. In the Griffiths phase we need to use the
large-argument limit of the scaling function giving $\chi \sim T^{d/z-1} r^{d\nu +
2\nu\psi(1-\phi)} \ln^2(\Omega_I/T)$. Thus, $\chi$ shows the nonuniversal power-law
temperature dependence characteristic of a quantum Griffiths phase. For $z>d$, the
susceptibility actually diverges with $T\to 0$.
Along the same lines, the scaling form of the local susceptibility is found to be
\begin{eqnarray}
\chi^{\rm loc}(r,T) &=& \frac{1}{T}\left[\ln(\Omega_I/T)\right]^{\phi-d/\psi}\Theta_{\chi}^{\rm
loc}\left[r^{\nu\psi}\ln(\Omega_I/T)\right]~,~~ \nonumber \\~
\label{eq:chi-scaling_loc}
\end{eqnarray}
with $\Theta^{\rm loc}_\chi(y)=X_n(y) X_\mu(y)$. This reduces to  $\chi^{\rm loc} \sim
\left[\ln(\Omega_I/T)\right]^{\phi-d/\psi}/T$ at criticality and to
$\chi^{\rm loc} \sim T^{d/z-1} r^{d\nu + \nu\psi(1-\phi)} \ln(\Omega_I/T)$ in the disordered
Griffiths phase.

\begin{figure}
\begin{center}\includegraphics[%
  clip,
  width=0.85\columnwidth]{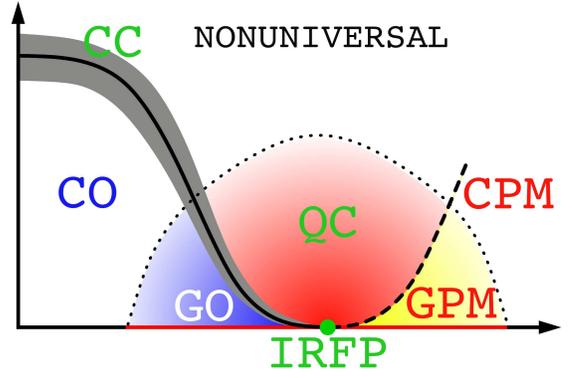}
\end{center}
\caption{(Color online) Temperature--coupling phase diagram for Ohmic dissipation. IRFP
denotes the infinite-randomness critical point. The ordered phase is divided into
a conventional (CO) region and a quantum Griffiths (GO) region. On the disordered side of
the transition, there is a quantum Griffiths paramagnet (GPM) followed by a conventional
quantum paramagnet (CPM). The phase boundary (solid) and the
crossover line (dashed) between the quantum critical (QC) region and the quantum
paramagnetic regions take unusual exponential forms leading to a wide quantum critical region.
In the classical critical region (CC) close to the phase boundary classical thermal
fluctuations dominate. At sufficiently high temperatures (above the dotted dome),
the behavior is nonuniversal.
\label{cap:phase-diagram}}
\end{figure}

The scaling form (\ref{eq:chi-scaling}) of the susceptibility can also be used to infer
the shape of the phase boundary close to the quantum phase transition. The
finite-temperature transition corresponds to a singularity in $\Theta_{\chi}(y)$ at some
nonzero argument $y_{c}$. This yields the unusual form $T_{c}\sim\exp(-{\rm
const}\left|r\right|^{-\nu\psi})$ shown in Fig.\ \ref{cap:phase-diagram}. The crossover
line between the quantum critical and quantum paramagnetic regions displays analogous
behavior.

The specific heat $C$ can be found by first adding the total energy contributions of all
surviving clusters,
\begin{eqnarray}
\Delta E(r,T) &=& T  n_T(r)  \nonumber \\
              &=& T \left[\ln(\Omega_I/T)\right]^{-d/\psi}
              \Theta_E\left[r^{\nu\psi}\ln(\Omega_I/T)\right]~,~~~
\label{eq:E-scaling}
\end{eqnarray}
with $\Theta_E(y)=X_n(y)$. After taking the temperature derivative this gives
$C \sim \left[\ln(\Omega_I/T)\right]^{-d/\psi}$ at criticality and $C \sim r^{d\nu}
T^{d/z}$ in the Griffiths phase.

To calculate the dependence of the low-temperature order parameter $m$ on an external
(conjugate) field $H$, we run the RG to the energy scale $\Omega_H =\mu_\Omega H \gg T$. All
decimated clusters have $\epsilon \gg \mu H$ and do not contribute significantly to the
order parameter. All surviving clusters have $\epsilon \ll \mu H$ and are fully
polarized. Summing over all surviving clusters therefore gives
\begin{eqnarray}
m(r,H) &=&  n_{\Omega_H}(r) \mu_{\Omega_H}(r) \nonumber\\
          &=&
          \left[\ln(\Omega_I/\Omega_H)\right]^{\phi-d/\psi}\Theta_{m}\left[r^{\nu\psi}
          \ln(\Omega_I/\Omega_H)\right]~,~~
\label{eq:m-scaling}
\end{eqnarray}
where $\Theta_m(y) = X_n(y) X_\mu(y)$. After resolving the implicit field dependence
caused by the moment in the definition of the energy scale $\Omega_H =\mu_\Omega H$, we
find $m \sim [\ln(\Omega_I/H)]^{\phi - d/\psi}$ (with double-logarithmic
corrections) at criticality, $r=0$. For $r>0$, we obtain a nonuniversal power-law field
dependence,
$m \sim  H^{d/z} r^{d\nu +\nu\psi(1-\phi)(1+d/z)}\left[\ln(\Omega/H)\right]^{1+d/z}$,
characteristic of a quantum Griffiths phase.

Finally, to find the zero-temperature dynamic susceptibility $\chi$ at external frequency $\omega$, we run
the RG to the energy scale $\Omega_\omega = \gamma \omega = \gamma_0 \mu_\Omega \omega$
($\gamma = \mu \gamma_0$ is the effective damping constant of a cluster of moment $\mu$).
All decimated clusters (having $\epsilon \gg \gamma \omega$) only make negligible contributions to $\chi$.
The surviving clusters have $\epsilon \ll \gamma \omega$ which simplifies
(\ref{eq:cluster-chi_dyn}) to ${\rm Im} \chi_{\rm cl}(\omega) = {\mu}/{\gamma_0 \omega}$. In
the same limit, the local dynamic susceptibility reads
${\rm Im} \chi^{\rm loc}_{\rm cl}(\omega) = 1/{\gamma_0 \omega}$. Using
(\ref{eq:n_scaling}) and (\ref{eq:mu_scaling}) we now sum over all surviving clusters to
obtain the scaling form
\begin{eqnarray}
\lefteqn{ {\rm Im} \chi(r,\omega) =} \nonumber \\
&=&\frac 1 {\gamma_0 \omega} \left[ \ln(\Omega_I/\Omega_\omega)
\right]^{\phi-d/\psi} \Theta_{\rm dyn}\left[r^{\nu\psi} \ln(\Omega_I/\Omega_\omega)
\right]~,~~
\label{eq:chi_scaling_dyn}
\end{eqnarray}
with $\Theta_{\rm dyn}(y) = X_n(y) X_\mu(y)$.
After resolving the implicit frequency dependence brought about by the moment in the
definition of $\Omega_\omega$, the leading low-frequency behavior of the dynamic susceptibility
at criticality is
${\rm Im} \chi \sim [\ln(\Omega_I/\gamma_0\omega)]^{\phi-d/\psi}/(\gamma_0\omega)$.
In the disordered Griffiths phase, we obtain
${\rm Im} \chi \sim (\gamma_0\omega)^{d/z-1} r^{d\nu+\nu\psi(1-\phi)(1+d/z)}
[\ln(\Omega_I/\gamma_0\omega)]^{1+d/z}$. The differences from the random transverse-field Ising model
results \cite{DamleMotrunichHuse00,MotrunichDamleHuse01b} have two reasons
(i) the additional frequency dependence hidden in the effective
damping constant $\gamma$ and (ii) the difference between a Lorentzian spectrum in our case
and the $\delta$-function spectrum in the Ising case.
The local dynamic susceptibility can be found along the same lines, yielding the scaling
form
\begin{eqnarray}
\lefteqn{ {\rm Im} \chi^{\rm loc}(r,\omega) = } \nonumber \\
&=&\frac 1 {\gamma_0 \omega} \left[ \ln(\Omega_I/\Omega_\omega)
\right]^{-d/\psi} \Theta_{\rm dyn}^{\rm loc}\left[r^{\nu\psi} \ln(\Omega_I/\Omega_\omega)
\right] ~,~~
\label{eq:chi_scaling_dyn_loc}
\end{eqnarray}
with $\Theta_{\rm dyn}^{\rm loc}(y) = X_n(y)$. At criticality, this leads to
${\rm Im} \chi^{\rm loc} \sim [\ln(\Omega_I/\gamma_0\omega)]^{-d/\psi}/(\gamma_0\omega)$,
and in the disordered Griffiths phase, we get ${\rm Im} \chi^{\rm loc} \sim (\gamma_0\omega)^{d/z-1} r^{d\nu+\nu\psi(1-\phi)d/z}
[\ln(\Omega_I/\gamma_0\omega)]^{d/z}$.

\section{Generalizations}
\label{sec:Generalizations}

\subsection{Non-Ohmic Dissipation}

In this section we briefly discuss how our results change, if we replace the Ohmic
damping term ($z_0=2$) in the starting action (\ref{eq:clean-action}),
(\ref{eq:bare_Gamma}) with a non-Ohmic term ($z_0\ne 2$). We are interested in the range
$z_0=1$ to $\infty$; $z_0=1$ corresponds to undamped (dissipationless) dynamics,
$1<z_0<2$ is the so-called super-Ohmic case (damping qualitatively weaker than Ohmic
damping) and for $z_0>2$, the damping is sub-Ohmic (qualitatively stronger than Ohmic).

Let us first consider sub-Ohmic damping,  $z_0>2$. In this case, the crucial
observation is that a single cluster with sufficiently large damping constant can
undergo a freezing or localization transition independent of the bulk system.
In Sec.\ \ref{subsec:single_cluster}
we showed that this transition occurs when the damping constant $\gamma$ becomes larger than
$\gamma_c=\Lambda^{(z_0-2)/z_0} z_0/(\pi(z_0-2))$.
Within the strong-disorder RG, the damping constant $\gamma$ renormalizes additively.
Thus, even for very small bare
dissipation, sufficiently large and strongly damped clusters will be formed under the RG
(as long as $\mu \to \infty$ with $\Omega \to 0$).
Once they are formed, their quantum dynamics freezes.
Consequently, for $z_0>2$ the global quantum phase transition is destroyed by
smearing.\cite{Vojta03a}

In the super-Ohmic case, $1\le z_0 <2$, the behavior is less exotic.
To study this case, we repeat the derivation of
the strong disorder RG recursion relations described in Subsection \ref{subsec:recursions}
for $1\le z_0 <2$. As was already pointed out, the multiplicative form of the
recursion (\ref{eq:J-tilde}) for the interactions $J$ follows directly from the structure
of second order perturbation theory and does not depend on $z_0$. In contrast, the
recursion for the gaps $\epsilon$ does depend on the value of $z_0$. Repeating the
exact solution of the two-site cluster for the super-Ohmic case, we find
\begin{equation}
\tilde \epsilon_2^{-x} = \alpha\left[\epsilon_2^{-x} + \epsilon_3^{-x}\right]
\label{eq:e-tilde-super}
\end{equation}
instead of the multiplicative form (\ref{eq:e-tilde}). Here $x=(2-z_0)/z_0$ and $\alpha$
is a constant. This form also follows from the fact that the damping constants add,
$\tilde \gamma_2 =\gamma_2 +\gamma_3$, together with the power-law dependence $\epsilon
\sim \gamma^{z_0/(z_0-2)}=\gamma^{-1/x}$ of the single-cluster gap on the damping
constant derived in (\ref{eq:gap-super-Ohmic}). For undamped dynamics, $z_0=1$, eq.\
(\ref{eq:e-tilde-super}) reduces to the dirty boson result $1/\tilde \epsilon_2 =
1/\epsilon_2 + 1/\epsilon_3$ derived by Altman \emph{et al.}\cite{AKPR04} These authors
also solved the resulting flow equations for $z_0=1$ and found Kosterlitz-Thouless-like
flows.

While a full solution of the RG flow equations in the generic case $1<z_0<2$
remains a task for the future, the qualitative critical behavior can be inferred
from the recursion relation (\ref{eq:e-tilde-super}). As a result of the additive
form of (\ref{eq:e-tilde-super}), the local gaps $\epsilon$ are much more
weakly renormalized than the interactions which are governed by the multiplicative
recursion (\ref{eq:e-tilde}). Near criticality, the distribution
of the interactions $J$ thus becomes highly singular while that of
the gaps $\epsilon$ remains narrower. We therefore expect the critical point
not to be of infinite-randomness type but conventional with power-law scaling
$\tau\sim\xi^{z}$, although the dynamical exponent $z$ can become
arbitrarily large as $z_{0}\rightarrow2^{-}$. Similar behavior was
found at a percolation quantum phase transition.\cite{VojtaSchmalian05b}

\subsection{Generic $N>1$}

So far, all of our explicit calculations have been for the large-$N$ limit of the $O(N)$
order parameter field theory. In this subsection we show that the results do not change
qualitatively for all $N>1$, i.e., all continuous symmetry cases. In order to do so, we
reanalyze the recursion relations (\ref{eq:J-tilde}) and (\ref{eq:e-tilde}) for generic $N$
(the relations (\ref{eq:moment})  and (\ref{eq:J-tilde-cluster}) trivially carry over for all
$N$). For definiteness, we focus on the case of Ohmic dissipation.

As discussed above, the multiplicative form of the recursion (\ref{eq:J-tilde}) for the
interactions relies on the structure of second order perturbation theory only, it is thus
valid for all $N$ including the discrete Ising case. In contrast, the form of the
recursion (\ref{eq:e-tilde}), which describes how the local gap $\epsilon$ (i.e.,
distance from criticality) changes if two clusters are combined, potentially does depend
on $N$. To understand this dependence, we first look at the related problem of the
dependence of $\epsilon$ on the size (moment) of the cluster.

By invoking the quantum-to-classical mapping it was recently shown\cite{VojtaSchmalian05}
that the gap depends exponentially on the size, $\epsilon
\sim e^{-c \mu}$ (with $c$ being a constant) for all continuous symmetry cases $N>1$.
This follows from the fact that classical one-dimensional continuous-symmetry
$O(N)$ models with $1/r^{2}$ interaction are known to be exactly \emph{at}
their lower critical dimension\cite{Joyce69,Dyson69,Bruno01} implying an exponential
dependence of the correlation length on the coupling strength. Alternatively, one can
explicitly estimate the strength of the \emph{transverse} fluctuations in a putative ordered
phase and notice the logarithmic divergence of $\int_0^\Lambda d\omega /(\gamma \omega)$.
The exponential size (moment) dependence of the gap $\epsilon$ requires a multiplicative
structure of the recursion relation (\ref{eq:e-tilde}) for the merging of two clusters
because their moments simply add, see Eq.\ (\ref{eq:moment}). We thus conclude that this
multiplicative structure is valid for all continuous symmetry cases, $N>1$.

Consequently, for sufficiently broad disorder distributions, the complete set of recursion
relations  (\ref{eq:J-tilde}), (\ref{eq:moment}),
(\ref{eq:J-tilde-cluster}) and (\ref{eq:e-tilde}) is valid for all $N>1$, and with it the
resulting infinite-randomness scenario of Sec.\ \ref{subsec:IRFP}. Possible $N$-dependent
prefactors modify nonuniversal quantities only. An analogous conclusion was
drawn in the undamped case, $z_0=1$, in Refs.\ \onlinecite{AKPR04,BMSV06}.

The universal behavior of all continuous symmetry cases has to be contrasted with
the case of Ising symmetry, $N=1$. In the Ising case, the gap does \emph{not} depend
exponentially on the cluster size. Instead, for sufficiently large Ohmic dissipation, the
cluster dynamics freezes, i.e., it undergoes the localization transition of the
dissipative two-state system.\cite{LCDFGZ87} The resulting behavior of an Ising system
with Ohmic dissipation is thus very similar to that of a continuous-symmetry system
with sub-Ohmic dissipation (as discussed in the last subsection): Sufficiently large
clusters freeze independently from the rest of the system which leads to a destruction of
the global quantum phase transition by smearing. This behavior was predicted in
Ref.\ \onlinecite{Vojta03a} and recently confirmed by an analytical strong-disorder RG \cite{HoyosVojta08}
as well as numerical simulations.\cite{SchehrRieger06,SchehrRieger08}

\section{Conclusions}
\label{sec:Conclusions}

In summary, we have studied quantum phase transitions in systems with continuous-symmetry
$O(N)$ order parameters under the influence of both quenched disorder and dissipative dynamics.
To this end, we have applied a strong-disorder RG to the LGW order parameter field theory
of the transition. For Ohmic dissipation, we have found an exotic infinite-randomness critical
point in the same universality class as the random transverse-field Ising chain.
In the sub-Ohmic case, the quantum phase transition is destroyed by smearing, while super-Ohmic
damping (including the undamped case) leads to conventional behavior. These results must be
contrasted with the case of Ising symmetry for which an infinite-randomness critical point
occurs in the absence of damping\cite{Fisher92,Fisher95} while Ohmic dissipation
causes a smeared quantum phase transition.\cite{Vojta03a,HoyosVojta08}

All these different behaviors and their relations can be understood with the help of a general
classification \cite{VojtaSchmalian05,Vojta06} of phase transitions in the presence of weak disorder.
This classification is based on the effective dimensionality of the defects or, equivalently,
the rare regions: If finite-size regions are exactly at the lower critical dimension of the
problem, the critical point is of infinite-randomness type (accompanied by power-law
quantum Griffiths singularities). Here this applies to continuous-symmetry order
parameters with Ohmic dissipation as well as dissipationless Ising order parameters.
If the rare regions are below the lower critical dimension, the behavior is
conventional (continuous symmetry order parameters with super-Ohmic dissipation);
and if they are above the lower critical dimension, individual regions order (freeze)
independently, leading to a smeared transition (continuous symmetry order parameters with
sub-Ohmic damping or Ising systems with at least Ohmic damping).

It is worth noting that Del Maestro \emph{et al.}\cite{DelMaestroRosenowSachdev08} very
recently studied the large-$N$ action (\ref{eq:action}) in one dimension by numerically solving the
saddle-point equations. All their results are in beautiful agreement with our
predictions, i.e., they confirmed that the quantum critical point is of
infinite-randomness type and in the universality class of the random transverse-field
Ising model.

We now turn to potential experimental realizations of our theory. One application is the
Hertz-Millis theory \cite{Hertz76,Millis93} of the (incommensurate) itinerant
antiferromagnetic quantum phase transition. In this theory, the LGW free energy
(\ref{eq:clean-action}) is derived from a microscopic Hamiltonian of interacting
electrons by integrating out the fermionic degrees of freedom in favor of the order
parameter field $\varphi$. While this procedure involves integrating out soft (gapless)
particle-hole excitations and is thus potentially
dangerous,\cite{BelitzKirkpatrickVojta05} the resulting order parameter field theory of
the antiferromagnetic transition appears to be internally consistent and free of
additional singularities, at least in three dimensions. However, the applicability of the
theory to realistic systems is still a controversial question, in particular for the
much-studied heavy fermion compounds where several experimental results are in pronounced
disagreement with the theoretical predictions.\cite{Stewart01,Stewart06} Different
scenarios to explain the discrepancies are discussed in the literature (see Ref.\
\onlinecite{LRVW07} for a recent review), and one much-discussed reason are disorder
effects. \cite{MirandaDobrosavljevic05}

\begin{figure}
\begin{center}\includegraphics[%
  clip,
  width=0.85\columnwidth,
  keepaspectratio]{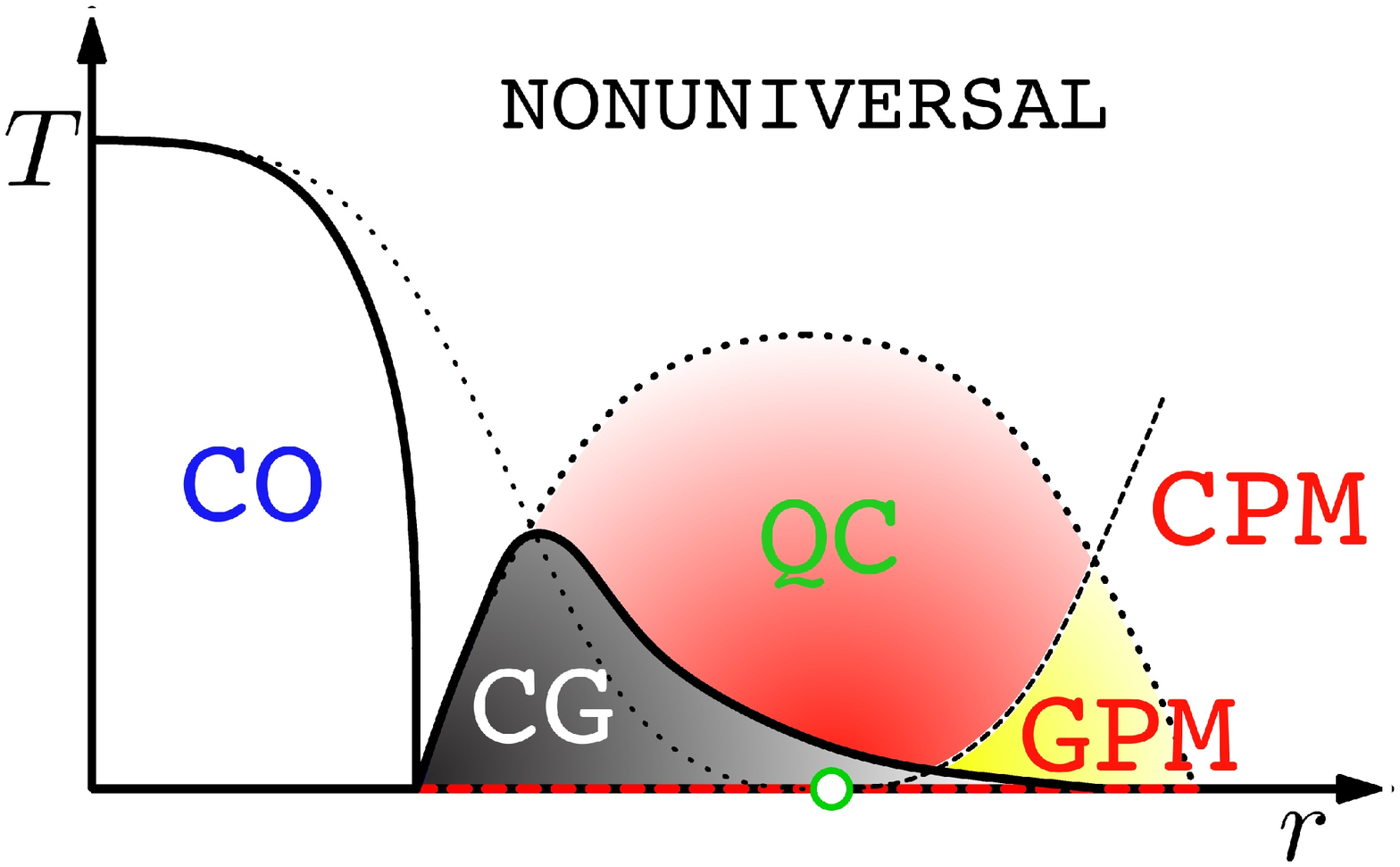}
\end{center}
\begin{center}\includegraphics[%
  clip,
  width=0.85\columnwidth,
  keepaspectratio]{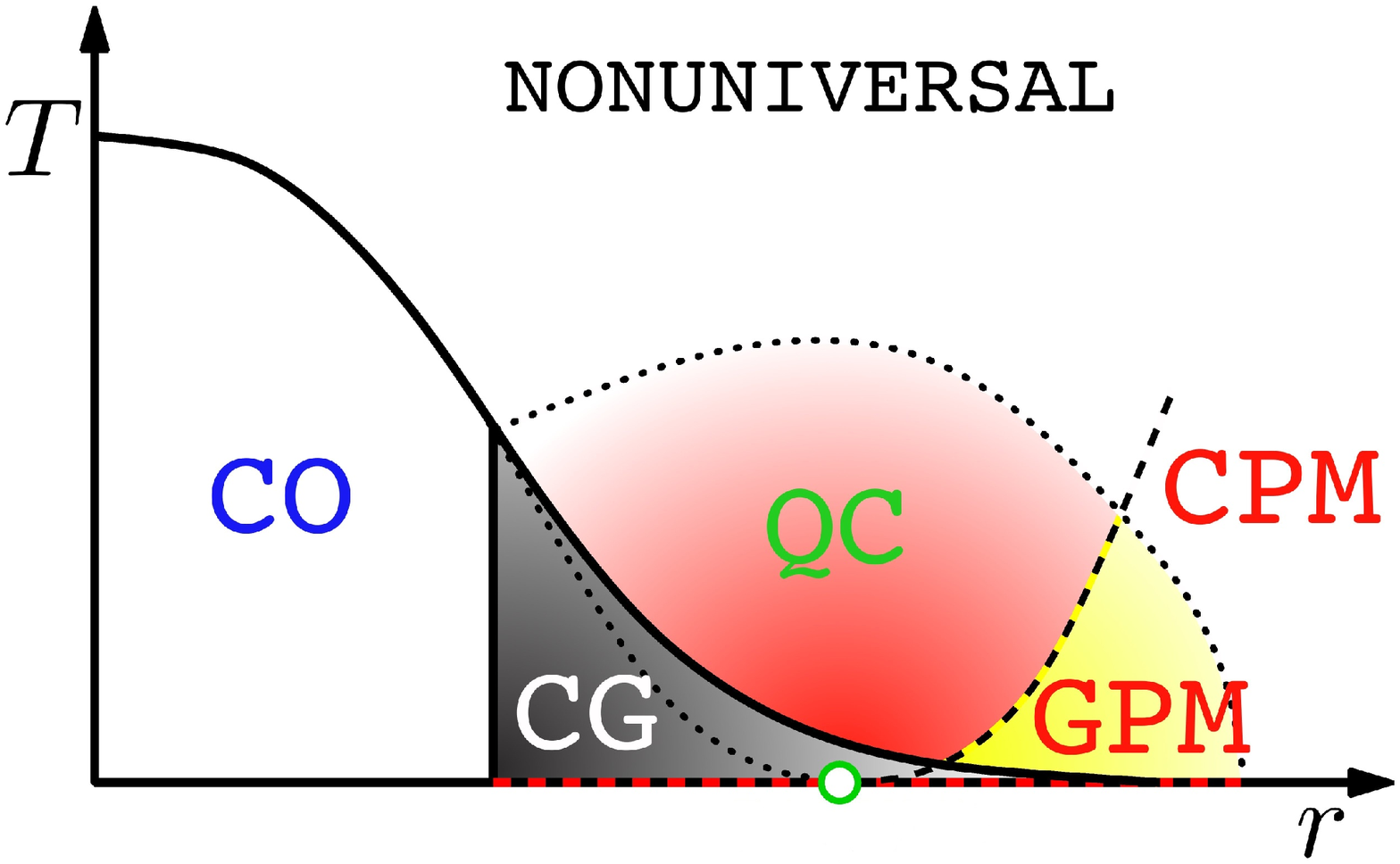}
\end{center}
\caption{(Color online) Schematic phase diagrams of a disordered itinerant quantum
antiferromagnet close to the quantum  phase transition showing two possible scenarios for
the appearance of a cluster glass phase, which is denoted by CG. The other labels are as
in Fig.\ \ref{cap:phase-diagram}.
\label{cap:CG}}
\end{figure}

Our theory provides explicit results on how the interplay of dissipation and disorder in
the vicinity of the itinerant antiferromagnetic quantum phase transition can yield
activated dynamics, quantum Griffiths phenomena, and non-Fermi liquid behavior. We expect
this to make an experimental verification or falsification of the disorder scenario much
easier. Note that a generic metallic system will have extra complications not contained
in the LGW free energy (\ref{eq:clean-action}). Specifically, attention must be paid to
the long-range Ruderman-Kittel-Kasuya-Yosida (RKKY) part of the interaction between the
magnetic fluctuations. It can produce an extra subohmic dissipation of locally ordered
clusters \cite{DobrosavljevicMiranda05} which leads to freezing into a ``cluster glass''
phase at a low non-universal temperature $T_{{\rm CG}}$ determined by the strength of the
RKKY interactions. This phase replaces part of the quantum Griffiths regions. It's
properties and the zero and finite-temperature transitions to the surrounding phases are
not fully explored, yet (the transitions may be of fluctuation-driven first order at low
temperatures\cite{CaseDobrosavljevic07}). The behavior of observables in the broad
quantum critical region above the cluster glass phase will be controlled by our
infinite-randomness critical point. Possible phase diagram scenarios are sketched in
Fig.\ \ref{cap:CG}.

Another potential application that has attracted considerable attention recently is the
superconductor-metal quantum phase transition occurring as a function of wire thickness
in ultrathin nanowires.\cite{RogachevBezryadin03} The clean version of this transition was studied by
means of a one-dimensional LGW theory (\ref{eq:clean-action}) with a complex order
parameter (equivalent to $N=2$) and Ohmic
dissipation.\cite{SachdevWernerTroyer04,DRSS08} However, there is experimental evidence
for the pair breaking in this system being caused by magnetic impurities at the surface
of the nanowire. This inevitably introduces quenched disorder due to the random positions
of the magnetic impurities. Our theory thus describes the thermodynamics of this quantum
phase transition. With proper modifications, it should also apply to arrays of resistively shunted
Josephson junctions.

So far, we have focused on the thermodynamics close to the quantum phase transition.
Transport properties can also be calculated within the strong-disorder RG by following
the approach of Refs.\ \onlinecite{DamleMotrunichHuse00,MotrunichDamleHuse01b}.
Calculations along these lines are underway; their results will be reported elsewhere.

\section*{Acknowledgements}

This work has been supported in part by the NSF under grant nos. DMR-0339147
and DMR-0506953, by Research
Corporation, and by the University of Missouri Research Board. We gratefully acknowledge
discussions with A. Del Maestro and S. Sachdev as well the hospitality of the Aspen
Center for Physics during part of this research.

\bibliographystyle{apsrev}
\bibliography{../../00Bibtex/rareregions}

\begin{thebibliography}{65}
\bibitem[{Bib()}]{Bibnote1}
\bibinfo{note}{In principle, one must distinguish between the bare distance
  from criticality appearing in (\ref {eq:bare_Gamma}) and the renormalized one
  appearing in the scaling relations. We will suppress this difference unless
  it is of importance for our considerations.}
\end{thebibliography}
\end{document}